\documentclass[conference]{IEEEtran}
\IEEEoverridecommandlockouts
\usepackage{cite}
\usepackage{amsmath,amssymb,amsfonts}
\usepackage{algorithmic}
\usepackage{graphicx}
\usepackage{textcomp}
\usepackage{xcolor}
\usepackage{comment}
\def\BibTeX{{\rm B\kern-.05em{\sc i\kern-.025em b}\kern-.08em
    T\kern-.1667em\lower.7ex\hbox{E}\kern-.125emX}}

\usepackage{amsthm}        
\usepackage{algorithm}     
\usepackage{booktabs}      
\usepackage{xspace}        
\usepackage{pifont}        
\usepackage{url}           
\usepackage{microtype}     

\newcommand{\cmark}{\ding{51}}
\newcommand{\xmark}{\ding{55}}

\newcommand{\AUCCten}{0.982}             
\newcommand{\TPRtenthCten}{0.753}        
\newcommand{\TPRhundredthCten}{0.629}    
\newcommand{\AUCChund}{0.975}            
\newcommand{\TPRtenthChund}{0.729}       
\newcommand{\TPRhundredthChund}{0.594}   
\newcommand{\AUCStl}{0.999}              
\newcommand{\TPRtenthStl}{0.993}         
\newcommand{\TPRhundredthStl}{0.987}     
\newcommand{\AUCCeleb}{0.879}            
\newcommand{\TPRtenthCeleb}{0.853}       
\newcommand{\TPRhundredthCeleb}{0.785}   

\newcommand{\AUCLoraSm}{0.991}           
\newcommand{\TPRtenthLoraSm}{0.857}      
\newcommand{\TPRhundredthLoraSm}{0.793}  
\newcommand{\AUCLoraLg}{0.996}           
\newcommand{\TPRtenthLoraLg}{0.911}      
\newcommand{\TPRhundredthLoraLg}{0.838}  
\newcommand{\AUCFullFT}{0.995}           
\newcommand{\TPRtenthFullFT}{0.879}      
\newcommand{\TPRhundredthFullFT}{0.799}  

\newcommand{\markLoraSmCross}{\cmark}   
\newcommand{\dCtwoLoraSmCross}{2.49} 
\newcommand{\markLoraSmRand}{\cmark}    
\newcommand{\dCtwoLoraSmRand}{19.0}  
\newcommand{\markLoraLgCross}{\cmark}   
\newcommand{\dCtwoLoraLgCross}{2.65} 
\newcommand{\markLoraLgRand}{\cmark}    
\newcommand{\dCtwoLoraLgRand}{19.1}  
\newcommand{\markFullFTCross}{\cmark}   
\newcommand{\dCtwoFullFTCross}{2.44} 
\newcommand{\markFullFTRand}{\cmark}    
\newcommand{\dCtwoFullFTRand}{19.1}  

\newcommand{\AUCMmd}{0.982}              
\newcommand{\TPRtenthMmd}{0.755}         
\newcommand{\TPRhundredthMmd}{0.629}     
\newcommand{\AUCSgd}{0.981}              
\newcommand{\TPRtenthSgd}{0.754}         
\newcommand{\TPRhundredthSgd}{0.623}     
\newcommand{\AUCNoiseSm}{0.981}          
\newcommand{\TPRtenthNoiseSm}{0.744}     
\newcommand{\TPRhundredthNoiseSm}{0.614} 
\newcommand{\AUCNoiseLg}{0.976}          
\newcommand{\TPRtenthNoiseLg}{0.796}     
\newcommand{\TPRhundredthNoiseLg}{0.684} 
\newcommand{\AUCDistill}{0.492}          
\newcommand{\TPRtenthDistill}{0.000}     
\newcommand{\TPRhundredthDistill}{0.000} 

\newtheorem{definition}{Definition}

\begin{document}

\title{Membership is Ownership: A Robust Ownership Verification Framework for Diffusion Models
\thanks{Corresponding Author: Zuobin Xiong}
\thanks{This paper has been accepted to IEEE International Conference on Data Mining (ICDM) 2026}
}

\author{\IEEEauthorblockN{
		Feng Jiang\IEEEauthorrefmark{1},
		Zuobin Xiong\IEEEauthorrefmark{2},
		An Huang\IEEEauthorrefmark{2},
        Zhipeng Cai\IEEEauthorrefmark{1}, and
		Yingshu Li\IEEEauthorrefmark{1}}
		\IEEEauthorblockA{
		\IEEEauthorrefmark{1}Department of Computer Science, Georgia State University, Atlanta, USA\\
		\IEEEauthorrefmark{2}Department of Computer Science, University of Nevada Las Vegas, Las Vegas, USA\\
		\IEEEauthorrefmark{1} {\it fjiang4@student.gsu.edu, \{yili, zcai\}@gsu.edu};
		\IEEEauthorrefmark{2} {\it \{zuobin.xiong, an.huang\}@unlv.edu}}
		}

\maketitle

\begin{abstract}
Large-scale diffusion models have fueled numerous profitable downstream applications for AI-related businesses, including visual editing and content creation.
Meanwhile, due to the huge amount of resource consumption (e.g., computation and high-quality data) during training, such diffusion models are deemed valuable intellectual property (IP) for tech companies like OpenAI and Google.
Yet, the IP assets are vulnerable to various unauthorized uses by adversaries seeking to steal models for customized, usually commercial applications.
Some existing approaches have explored IP protection for AI models; however, they mostly face structural limitations in common --- using a training-time watermarking by injecting artifacts in the model, which can impose a measurable utility cost and can be weakened by post-hoc fine-tuning.
To address these challenges, this work investigates IP protection (i.e., model ownership verification) for diffusion models in a realistic commercial scenario with minimal model utility loss. 
Specifically, the proposed method builds a framework for model ownership verification, termed ``{Membership is Ownership} (MiO)'', based on a population-level hypothesis test on a private member evidence dataset.
MiO verifies ownership using two criteria: model attribution through
membership inference and model separation from public references.
Both are tested at $p<10^{-6}$.
We evaluate MiO on DDIM and Stable Diffusion models without modifying
the owner model or its sampling pipeline, and report ROC-AUC and
true-positive rates at fixed nominal false-positive targets.
Furthermore, MiO stays stable under different post-theft fine-tuning and weight perturbation in adversarial scenarios, reflecting better robustness compared to the watermarking methods.
\end{abstract}


\section{Introduction}
\label{sec:intro}

Large-scale generative diffusion models have become the critical infrastructure for image generation~\cite{rombach2022high}, producing numerous profitable downstream applications.
However, the impressive capability of these models is bought at considerable costs: the intensive computational resources and high-quality training data, which make the resulting generative AI models valuable intellectual property (IP) for the institutions and companies that own them.
Yet these IP assets are a vulnerable mining target to malicious theft, an adversary who aims to steal the model, which can then fine-tune and redeploy the model for other purposes without contribution, sabotaging the IP copyright and ownership.
Establishing \emph{model ownership verification} framework, therefore, is a pressing issue of practical economic and legal significance.

\emph{Ownership verification} is to assert whether a suspect model/checkpoint originates from a particular owner's (i.e., victim) base model.
Existing approaches to diffusion model IP protection address this problem along two complementary lines.
(1) \emph{Training-time watermarking} embeds an explicit signature into model parameters or output behavior~\cite{peng2023watermark,zhao2023recipe,wang2025sleepermark}, e.g., adding invisible noise into parameter space or output space for later identification, which are effective in some controlled settings, but imposes a measurable utility cost and can be progressively weakened by post-hoc fine-tuning, a common strategy used by adversaries.
(2) \emph{Data memorization verification} for diffusion models~\cite{kong2024efficient,matsumoto2023membership,hu2023loss} exploit the model's intrinsic memorization gap -- reconstruction error on training samples is systematically lower than on unseen samples -- and can verify used training data.
However, the memorization verification is fundamentally a per-sample classifier where a low reconstruction error on a \textbf{single sample} cannot certify \textbf{a model ownership} claim, because some individual samples may be easy to reconstruct.

The challenges above suggest another approach: conducting model ownership verification at the population level while minimizing impact on model utility.
In this paper, we present \textbf{Membership is Ownership} (MiO), a verification framework built on a principled observation: although the per-sample memorization signal is weak to claim ownership, the \emph{population-level} memorization pattern over a privately-curated evidence set is both robust and statistically auditable.
Particularly, the model trainer/owner reserves a private subset $\mathcal{W}$ of the training data, which is never disclosed to the public, and serves as a non-invasive watermark that leaves model parameters and output quality entirely intact.
To extract a robust signal from a suspect model for ownership verification, we introduce a multi-timestep $t$-error score via forward noising and single-step reconstruction by using different statistics (e.g., the 25th percentile) to suppress noise while retaining the memorization.
We use Gaussian quantile-regression networks to derive closed-form
thresholds for specified FPR targets under the conditional Gaussian
model.

We validate the framework across two evaluation scenarios in naive diffusion models, like DDIMs trained from scratch and commercialized diffusion models, like Stable Diffusion fine-tuned from the production-scale datasets.
For both scenarios, we report detection performance at fixed FPR targets.

In summary, the contributions of this work are as follows.
\begin{itemize}
    \item We reframe the model ownership verification into a population-level hypothesis test problem, which leads to a statistically auditable framework.

    \item The ownership verification process introduces minimal impact (i.e., 0 change in FID score) on the generation quality of original models by using a non-invasive evidence dataset.

    \item The proposed two-point verification protocol uses Gaussian
    quantile regression for attribution and Cohen's $d$ for reference
    separation, with a closed-form threshold for a chosen nominal
    per-sample false-positive rate under the conditional Gaussian model.

    \item We validate the framework on two scenarios across multiple datasets with different configurations, showing that MiO outperforms existing methods in verification accuracy, costs, utility loss, and post-theft robustness.
\end{itemize}

\section{Related Works}
\label{sec:related}

\textbf{Model Watermarking.}
Watermarking embeds an owner-specific signal that can be recovered later to prove provenance. 
The methods differ mainly in where the signal lives, which decides how hard it is to remove.
One line places it in the generation process. 
Gaussian Shading~\cite{yang2024gaussian} hides the message in the initial latent while keeping its distribution Gaussian, so training and image quality are less affected. 
Newer variants, Gaussian Shading++~\cite{yang2025gaussianshadingpp}, turn repeated latent embeddings into a per-bit likelihood ratio for soft decoding, the pseudorandom-code watermark of Gunn et al.~\cite{gunn2025undetectable} makes the mark cryptographically undetectable in the sampled noise, ROBIN~\cite{huang2024robin} moves the mark to an intermediate denoising state, and GaussMarker~\cite{li2025gaussmarker} adds a spatial-domain channel. 
Because the signal rides on the seed and the sampling path, a user who controls generation can drop it by resampling the initial latent.

A second line binds the signal to the model. 
The watermark diffusion process~\cite{peng2023watermark} and the trigger-based recipe of Zhao et al.~\cite{zhao2023recipe} install it at training time; 
Stable Signature~\cite{fernandez2023stable} fine-tunes the latent decoder so every output carries a fixed bitstring; ProMark~\cite{asnani2024promark} marks the training images and trains the model to reproduce them. SleeperMark~\cite{wang2025sleepermark} carries this to diffusion models by separating watermark knowledge from semantic knowledge during training, so the trigger is not forgotten when the stolen model is fine-tuned for a new task. 
Those model-related marks resist resampling, but they cost measurable utility and can be mitigated by removal attacks that weaken such watermarks while preserving image quality~\cite{zhao2024invisible,hu2025transfer,shamshad2025erasing,alam2025sadre,duan2025visual,pautov2024probabilistically}.

\textbf{Data Memorization Verification.}
The memorization of training data in model parameters can be used to check model or data attribution~\cite{maini2021dataset,dubinski2024cdi}.
Specifically, membership inference attack (MIA), an attack to decide whether an input is part of a model's training data~\cite{shokri2017membership,yeom2018privacy,salem2019ml}, has been used in literature to verify models.
For diffusion models, recent works investigate MIA using reconstruction-based signals~\cite{carlini2023extracting,duan2023diffusion,kong2024efficient}, loss/likelihood-based criteria~\cite{matsumoto2023membership,hu2023loss}, or scalable quantile-regression formulations~\cite{bertran2023scalable,tang2024membership}.
These methods are primarily evaluated as per-example membership decision (i.e., a binary classification of member or non-member) and are typically reported via attack metrics.
Ownership verification, however, requires comparative, population-level evidence: the ownership claim must be supported by the aggregate evidence on a designated evidence set via comparison with calibrated reference models to show separation, and tested against the suspect model via a statistically interpretable hypothesis rather than single-sample predictions.
One recent work, CDI~\cite{dubinski2024cdi}, adopted a similar idea to ours by aggregating several per-sample MIA, but they focused on data ownership, not the model ownership. 
MiO differs by curating a private evidence set as the ownership anchor
and applying a closed-form threshold for a specified FPR target before
the population test.


\begin{figure*}[t]
    \centering
    \includegraphics[width=1\linewidth]{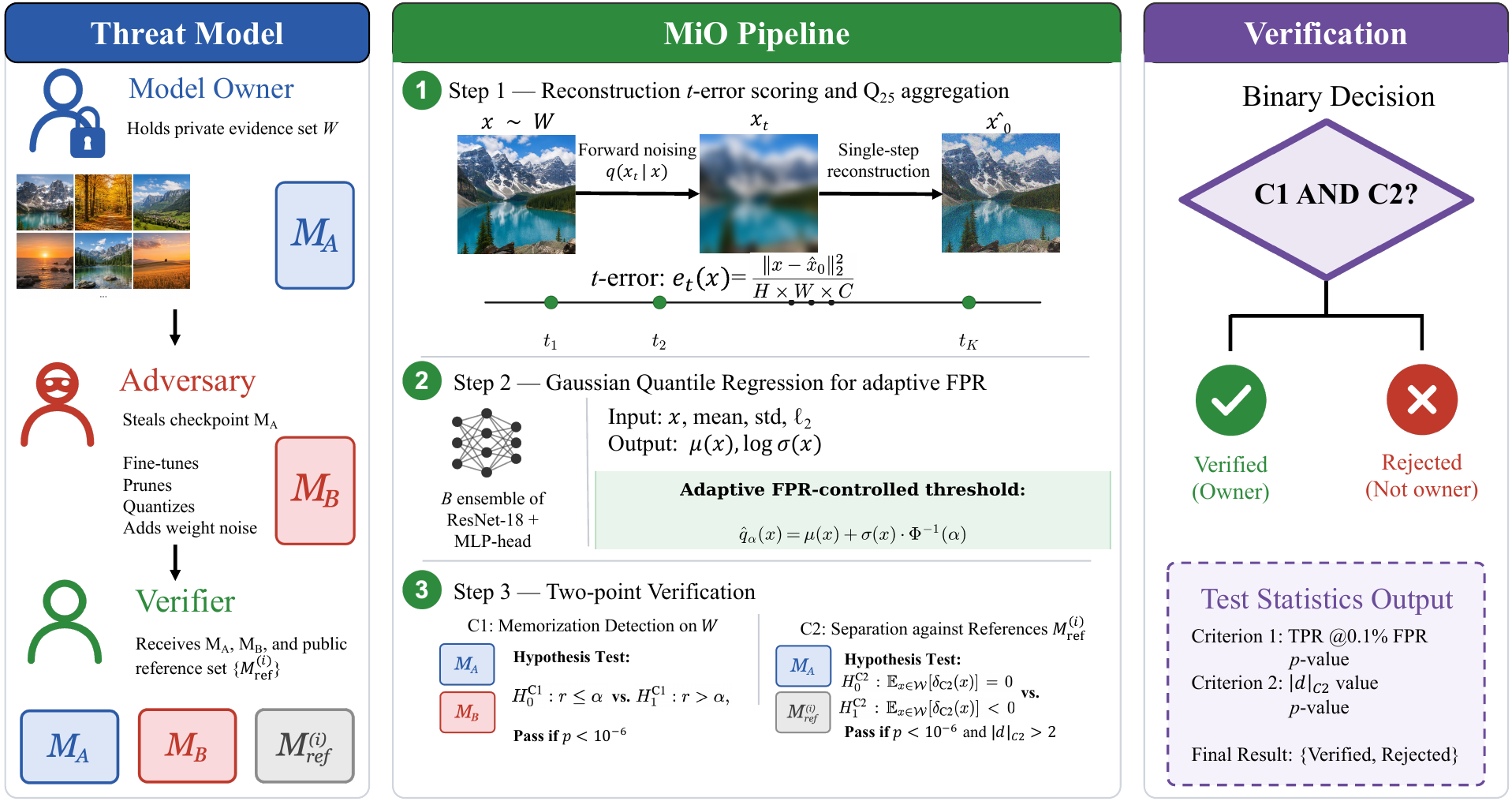}
    \caption{The framework and operation steps in the proposed MiO method.}
    \label{fig:placeholder}
\end{figure*}

\section{Problem Formulation}
\label{sec:problem}

\subsection{Problem Statement}

\subsubsection{Diffusion Models.}
A diffusion model corrupts a clean sample $x$ via a forward
noising process,
\begin{equation}
\label{eq:dm}
    x_t = \sqrt{\bar\alpha_t}\, x
        + \sqrt{1-\bar\alpha_t}\,\epsilon,
    \quad \epsilon \sim \mathcal{N}(0, I),
\end{equation}
where $\bar\alpha_t = \prod_{s=1}^{t}\alpha_s$, $x_t$ is the noised image at timestep $t \in \{1,\ldots,T\}$ and trains a
network $\epsilon_\theta(x_t, t)$ to predict the added noise.
The network $\epsilon_\theta$ is optimized by minimizing the denoising objective with $t$ drawn uniformly from $\{1,\ldots,T\}$. 
\begin{equation}
\label{eq:dm_loss}
    \mathcal{L}(\theta) = \mathbb{E}_{x,\,t,\,\epsilon}
        \big\| \epsilon - \epsilon_\theta(x_t, t) \big\|_2^2,
\end{equation}

New samples are produced by iteratively denoising from $x_T \sim \mathcal{N}(0, I)$; we adopt the deterministic DDIM sampler, which drops the stochastic term and admits a closed-form estimate $\hat{x}_0$ of the clean signal from any $(x_t, t)$ through $\epsilon_\theta$.
This $\hat{x}_0$ estimate is precisely the quantity on which our reconstruction score $e_t$ is built (Section~\ref{sec:t_error}).

Latent diffusion models (LDMs), e.g.\ Stable Diffusion~\cite{rombach2022high}, run the diffusion process in a compressed latent space, where a pretrained autoencoder maps an image $x$ to $z = \mathcal{E}(x)$ and decodes it as $\mathcal{D}(z)$.
The forward process and training objective (Eqs.~\eqref{eq:dm}--\eqref{eq:dm_loss}) then act on $z$ in place of $x$, and the denoiser $\epsilon_\theta(z_t, t, c)$ is conditioned on a text embedding $c$ injected through the U-Net cross-attention layers, which are the same layers LoRA adapters modify.
For LDMs, the reconstruction error $e_t$ can be evaluated in either latent space or pixel space, and the ablation study is evaluated in Section~\ref{sec:ablation}.

Both diffusion models systematically memorize portions of their training data~\cite{carlini2023extracting, somepalli2023diffusion} -- a memorization gap our framework exploits in Section~\ref{sec:method}.

\subsubsection{Membership Inference vs. Ownership Verification}
MIA leverages the memorization gap to determine whether a single query sample belongs to the training set of a target model~\cite{duan2023diffusion, kong2024efficient}, while the ownership verification is different.
We state this problem formally:

\begin{definition}[Membership Inference Attack]
Given a target model $M$ trained from $\mathcal{D}_{\mathrm{train}}$ and a query sample $x$,  MIA is a binary classifier $\mathcal{A}: (M, x) \rightarrow \{0, 1\}$ that predicts whether $x \in \mathcal{D}_{\mathrm{train}}$.
\end{definition}

However, ownership verification requires a population-level test:
significant memorization on a \emph{designated evidence set} relative to independent reference models.

\begin{definition}[Model Ownership Verification]
Let $\mathcal{M}_A$ be the owner's base model, $\mathcal{M}_B$ be a suspect model to be verified, and $\mathcal{W} \subset \mathcal{D}_{\mathrm{train}}(\mathcal{M}_A)$ be a private evidence set.
An ownership verification protocol is a hypothesis test that decides whether $\mathcal{M}_B$ is stolen from $\mathcal{M}_A$ under a controlled false-positive rate $\alpha$.
$\mathcal{V}: (\mathcal{M}_B, \mathcal{W}, \{M_{\text{ref}}^{(i)}\})
\rightarrow \{\textnormal{\textsc{Verified}}, \textnormal{\textsc{Rejected}}\}$  

\end{definition}

\textbf{Reference Models.}
The reference model set $\{\mathcal{M}_{\text{ref}}^{(i)}\}_{i=1}^{|\mathcal{R}|}$ in the definition above should satisfy three conditions: (i)~its training data is disjoint from the private set $\mathcal{W}$;
(ii)~it has independent provenance from $\mathcal{M}_A$;
and (iii)~it is type-compatible with the $t$-error computation procedure of Section~\ref{sec:t_error} -- i.e., a diffusion model on which a per-sample reconstruction score on $\mathcal{W}$ is well-defined.

\subsection{Threat Model}
\label{sec:threatmodel}
We consider three parties in the verification framework: the model \textbf{owner} who trained $\mathcal{M}_A$ and holds $\mathcal{W}$, the \textbf{adversary} $A$ who obtained $\mathcal{M}_A$ through theft (e.g., reverse engineering attacks or insider attacks) and produced a derivative model $\mathcal{M}_B$ via fine-tuning, and the \textbf{verifier} $V$ who judges whether $\mathcal{M}_B$ is derived from $\mathcal{M}_A$.

We assume the adversary $A$ has white-box access to $\mathcal{M}_A$ (due to theft) and may possess similar domain data disjoint from $\mathcal{W}$.
The adversary can apply adaptations, such as fine-tuning, weight perturbation, quantization, or pruning, to the stolen model weights to escape from the verification protocol.
Our threat model focuses on adaptations of the stolen model weights and
excludes independent retraining from scratch.
Nonetheless, we still examine distillation as a threat-model boundary case in Section~\ref{sec:theft} (Table~\ref{tab:distillation}).
The verifier $V$ (i.e., copyright judge or court) has white-box access to both $\mathcal{M}_A$, $\mathcal{M}_B$, and a set of public reference models $\{\mathcal{M}_{\text{ref}}^{(i)}\}_{i=1}^{|\mathcal{R}|}$, representing the null hypothesis.

The verification framework is established on the core fact that the adversary cannot access the owner's private evidence set $\mathcal{W}$, which should be the confidential information of a company.
In the worst case, even if $\mathcal{D}_{\mathrm{train}}$ were leaked, selecting the exact subset $\mathcal{W}$ from
training samples $\mathcal{D}_{\mathrm{train}}$ is computationally impossible with probability at most $P(\text{exact match})
    = \frac{1}{\binom{N}{|\mathcal{W}|}}
    \;\leq\; \left(\frac{|\mathcal{W}|}{N}\right)^{|\mathcal{W}|}$, which is $0.1^{5000}$ under the small CIFAR-10 setting
($N=50{,}000$, $|\mathcal{W}|=5{,}000$).

\section{MiO: Ownership Verification Framework}
\label{sec:method}
The MiO is a three-step verification framework that combines the following parts:
(1) a reconstruction error $e_t(x)$ to quantify model memorization on each timestep $t$;
(2) a Gaussian quantile regression-based function to derive a closed-form member distribution at any chosen false positive level;
and 
(3) a two-point verification criterion for ownership verification.
The framework of MiO is described in Fig.~\ref{fig:placeholder}.

\subsection{Reconstruction $t$-error Scoring}
\label{sec:t_error}
Diffusion models reconstruct their training data with lower error than unseen data~\cite{carlini2023extracting} and thus, the \emph{memorization gap} can be used to quantify membership.
Specifically, given a clean sample $x$, we first corrupt it via the forward diffusion schedule as shown in Eq.~\eqref{eq:dm} to obtain $x_t$.
The model then predicts the added noise $\hat{\epsilon}_\theta(x_t, t)$, from which we can recover ${x}_0$ from ${x}_t$ via
$\hat{x}_0 = \frac{x_t - \sqrt{1 - \bar{\alpha}_t} \cdot \hat{\epsilon}_\theta(x_t, t)}{\sqrt{\bar{\alpha}_t}}$.

We define the reconstruction \emph{$t$-error}, $e_t(x)$, as the pixel-level mean squared error at timestep $t$ between the original and reconstructed data as follows:
\begin{equation}
e_t(x) = \frac{\|x - \hat{x}_0\|_2^2}{H \times W \times C}
\end{equation}
where $H$, $W$, and $C$ denote the height, width, and number of channels, respectively.
For latent diffusion models such as Stable Diffusion, the reconstruction error calculation is applied in the VAE latent space without modification.
The empirical pixel-vs-latent comparison is in Section~\ref{sec:ablation}.
The full procedure is summarized in Algorithm~\ref{alg:t_error},  distinct from the single training timestep $t$ in Eq.~\eqref{eq:dm_loss}, the timestep set $\mathcal{T}=\{t_1,\ldots,t_K\}$ collects $K$ timesteps drawn independently and uniformly from $\{1,\ldots,T\}$.

\begin{algorithm}[t]
\caption{$t$-error Computation}
\label{alg:t_error}
\begin{algorithmic}
\REQUIRE Diffusion model ${\color{black}\epsilon_\theta}$, sample $x {\color{black}\in \mathbb{R}^{H \times W \times C}}$, timesteps $\mathcal{T} = \{t_1, \ldots, t_K\}$, noise schedule $\{\bar{\alpha}_t\}$
\ENSURE Aggregated score $s(x)$
\STATE $\texttt{errors} \leftarrow []$
\FOR{$t \in \mathcal{T}$}
    \STATE $\epsilon \sim \mathcal{N}(0, I)$
    \STATE $x_t \leftarrow \sqrt{\bar{\alpha}_t} \cdot x + \sqrt{1 - \bar{\alpha}_t} \cdot \epsilon$
    \STATE $\hat{\epsilon} \leftarrow {\color{black}\epsilon_\theta}(x_t, t)$
    \STATE $\hat{x}_0 \leftarrow (x_t - \sqrt{1 - \bar{\alpha}_t} \cdot \hat{\epsilon}) / \sqrt{\bar{\alpha}_t}$
    \STATE $e_t \leftarrow \|x - \hat{x}_0\|_2^2 / (H \times W \times C)$
    \STATE $\texttt{errors}.\text{append}(e_t)$
\ENDFOR
\STATE \textbf{return} $s(x) \leftarrow {Q_{25}}(\texttt{errors})$
\end{algorithmic}
\end{algorithm}

\subsection{Gaussian {Quantile Regression} for Adaptive Threshold}
\label{sec:score}

A single timestep $t$-error provides a noisy estimate of reconstruction quality, since the magnitude of $e_t(x)$ varies substantially with $t$.
To obtain a more stable statistic $s(x)$, we aggregate $t$-errors over $K$ uniformly sampled timesteps via different percentiles $Q_d$.
That is
\begin{equation}
  s(x) = Q_{d}\!\bigl(\{e_{t_i}(x)\}_{i=1}^{K}\bigr).
\end{equation}
We use $K = 50$ in the experiments, and an ablation study in Section~\ref{sec:ablation} shows that $K \geq 25$ will produce a stable regime.
For the percentile $Q_d$, a low percentile is preferred because the memorization gap between members and non-members is more pronounced at the timesteps where reconstruction is more accurate on members.
The 25th percentile ($Q_{25}$) is selected based on our empirical evaluation in Section~\ref{sec:ablation}, which retains the best performance while suppressing the upper tail.

We therefore model the conditional distribution of $s(x)$ for adaptive threshold construction.
Concretely, we apply a log transform to the score and parameterize the result as a conditional Gaussian distribution whose parameters are learned from $x$:
\begin{equation}
  y = \log(1 + s(x)) \sim \mathcal{N}\!\bigl(\mu(x),\, \sigma^2(x)\bigr).
\end{equation}

The conditional Gaussian parameters are produced by a learned predictor $f_\psi$ with parameters $\psi$ that takes sample $x$ and a 3-dimensional summary $(\mathrm{mean},\,\mathrm{std},\,\ell_2)$ of the $t$-error sequence $\{e_{t_i}(x)\}_{i=1}^{K}$, and outputs the mean and the \emph{log} standard deviation, $f_\psi: (x, summary) \mapsto (\mu(x),\,\log\sigma(x))$.
Predicting $\log\sigma$ rather than $\sigma$ leaves the head output unconstrained while guaranteeing a positive scale and avoiding the $\sigma\!\to\!0$ blow-up of the likelihood.
We then recover $\sigma(x) = \exp(\log\sigma(x))$.  
In the experiment design, the Gaussian learner $f_\psi$ is a ResNet-18 backbone, and the feature is concatenated and passed through an MLP head.
$f_{\psi}$ is trained on an auxiliary set $\mathcal{D}_{\mathrm{aux}}$, drawn from the same domain as $\mathcal{M}_A$'s training data but \emph{disjoint}
from it, by minimizing the Gaussian negative log-likelihood shown in Eq.~\eqref{eq:gaussian_loss}.
This disjointness matters because $\mathcal{D}_{\mathrm{aux}}$ defines the \emph{non-member} score distribution that calibrates the threshold.
In practice, we use a held-out split of the same dataset or public data of the same domain and exclude both $\mathcal{W}$ and the rest of $\mathcal{D}_{\mathrm{train}}(\mathcal{M}_A)$.

\begin{equation}
\label{eq:gaussian_loss}
\mathcal{L}(\psi)
= \mathbb{E}_{x \sim \mathcal{D}_{\mathrm{aux}}}\!\left[
\tfrac{1}{2}\log\sigma^2(x)
+ \frac{\bigl(y - \mu(x)\bigr)^2}{2\sigma^2(x)}
\right]
\end{equation}
Members produce systematically lower scores $s(x)$ than non-members, so the membership signal lies in the \emph{lower tail} of the non-member score distribution. 
For a target FPR $\alpha$, we set the adaptive threshold
$\hat{q}_\alpha(x)$ to the modeled $\alpha$-quantile of
$\log(1+s(x))$ for non-members conditional on $x$.
Under the conditional Gaussian model, this quantile
admits the closed form
\begin{equation}
\hat{q}_\alpha(x) = \mu(x) + \sigma(x)\,\Phi^{-1}(\alpha),
\end{equation}
where $\Phi^{-1}$ is the standard-Gaussian quantile function.
The threshold gives a per-sample, FPR-calibrated instruction for membership claim: we flag $x$ as a
likely member if $\log(1+s(x)) \leq \hat{q}_\alpha(x)$, or
equivalently by the threshold:
\begin{equation}
s(x) \leq \exp\bigl(\hat{q}_\alpha(x)\bigr) - 1.
\end{equation}

To reduce the predictor variance, we replace the single $f_\psi$ with a
bagging ensemble of $B=50$ predictors trained on $80\%$ bootstrap
resamples of $\mathcal{D}_{\mathrm{aux}}$, and use the averaged
$\hat{q}_\alpha(x)$ at test time.

\begin{algorithm}[t]
\caption{Gaussian Quantile Regression Training}
\label{alg:gaussian_qr}
\begin{algorithmic}
\REQUIRE Auxiliary non-member data $\mathcal{D}_{\mathrm{aux}}$, ensemble size $B$, bootstrap ratio $r$, max epochs $E$
\ENSURE Trained ensemble $\{f_{\psi^{(b)}}\}_{b=1}^{B}$
\FOR{$b = 1$ to $B$}
    \STATE $\mathcal{D}_b \leftarrow \text{BootstrapSample}(\mathcal{D}_{\mathrm{aux}}, r)$ \COMMENT{$r=0.8$, with replacement}
    \STATE $\mathcal{D}_b^{\mathrm{fit}}, \mathcal{D}_b^{\mathrm{val}} \leftarrow \text{Split}(\mathcal{D}_b, 0.9)$
    \STATE Initialize $f_{\psi^{(b)}}$ with ResNet-18 backbone
    \FOR{epoch $= 1$ to $E$}
        \FOR{batch $(x, s(x), \phi(x))$ in $\mathcal{D}_b^{\mathrm{fit}}$}
            \STATE $y \leftarrow \log(1 + s(x))$
            \STATE $(\mu, \log\sigma) \leftarrow f_{\psi^{(b)}}(x, \phi(x))$
            \STATE $\sigma \leftarrow \exp(\log\sigma)$ \COMMENT{positive scale}
            \STATE $\mathcal{L} \leftarrow \tfrac{1}{2}\log\sigma^2 + (y - \mu)^2 / (2\sigma^2)$
            \STATE Update $\psi^{(b)}$ by gradient descent on $\mathcal{L}$
        \ENDFOR
        \IF{validation loss on $\mathcal{D}_b^{\mathrm{val}}$ has not improved for $10$ epochs}
            \STATE \textbf{break} \COMMENT{early stopping}
        \ENDIF
    \ENDFOR
\ENDFOR
\STATE \textbf{return} $\{f_{\psi^{(b)}}\}_{b=1}^{B}$
\end{algorithmic}
\end{algorithm}
\subsection{Two-Point Verification Criteria}
\label{sec:criteria}

Built on the quantile score $s(x)$ from Section~\ref{sec:score}, this section defines the model ownership verification criterion.
Given the owner model $\mathcal{M}_A$, the suspect model $\mathcal{M}_B$, a set of public references $\{\mathcal{M}_{\text{ref}}^{(i)}\}$, and the private evidence set $\mathcal{W}$, the following criteria must be satisfied to declare verified ownership.

\textbf{Criterion 1: Memorization Detection on $\mathcal{W}$.}
For the target FPR $\alpha$, we apply the per-sample threshold
$\hat{q}_\alpha(x)$ from Section~\ref{sec:score} to each
$x \in \mathcal{W}$ under the suspect model $\mathcal{M}_B$ and compute
the empirical hit rate $\hat{r}$:
\begin{equation}
\label{eq:c1_rate}
\hat{r} = \frac{1}{|\mathcal{W}|} \sum_{x \in \mathcal{W}} \mathbf{1}\!\bigl[s_{\mathcal{M}_B}(x) \leq \exp(\hat{q}_\alpha(x)) - 1\bigr].
\end{equation}
Let $r$ denote the population hit rate $\Pr_{x \in \mathcal{W}}\!\bigl[s_{\mathcal{M}_B}(x)
\leq \exp(\hat{q}_\alpha(x))-1\bigr]$ that $\hat{r}$ estimates, and test the one-sided pair
\[
H_0^{\text{C1}}: r \leq \alpha \qquad \text{vs.} \qquad H_1^{\text{C1}}: r > \alpha,
\]
where the null hypothesis $H_0^{\text{C1}}$ states that $\mathcal{M}_B$ has no memorization of $\mathcal{W}$ beyond the calibrated false-positive rate, 
whereas $H_1^{\text{C1}}$ states that $\mathcal{M}_B$ retains memorization of $\mathcal{W}$ beyond it.
At the null boundary $r=\alpha$, the count $|\mathcal{W}|\cdot \hat{r}$ follows $\mathrm{Binomial}(|\mathcal{W}|, \alpha)$, so a one-sided exact binomial test can be applied.
The criterion~1 is declared passed when its $p$-value falls below $10^{-6}$.

\textbf{Criterion 2: Separation against Reference Models.}
The owner model $\mathcal{M}_A$ must reconstruct evidence-set samples significantly better than any public reference model, verifying the unique memorization of $\mathcal{M}_A$ on the private dataset $\mathcal{W}$.
For each reference model $\mathcal{M}_{\text{ref}}^{(i)}$, we first compute the per-sample difference
$\delta_{\text{C2}^{(i)} }(x) = s_{\mathcal{M}_A}(x) - s_{\mathcal{M}_{\text{ref}}^{(i)}}(x)$,
and aggregate it across the evidence set into Cohen's distance $d$ as follows:
\begin{equation}
\label{eq:paired_c2}
|d|_{\text{C2}} = \frac{|\,\text{mean}(\delta_{\text{C2}}(\mathcal{W}))\,|}{\text{std}(\delta_{\text{C2}}(\mathcal{W}))}.
\end{equation}
Then we test the null hypothesis $H_0^{\text{C2}}: \mathbb{E}_{x \in \mathcal{W}}[\delta_{\text{C2}}(x)] = 0$
against the one-sided alternative $H_1^{\text{C2}}: \mathbb{E}_{x \in \mathcal{W}}[\delta_{\text{C2}}(x)] < 0$.
{$H_0^{\text{C2}}$ says the owner reconstructs $\mathcal{W}$ no better than a public model that never saw it.
$H_1^{\text{C2}}$ says the owner's $t$-error on $\mathcal{W}$ is strictly lower than the reference's, a gap that only training on $\mathcal{W}$ can produce.}
Criterion~2 passes when a one-sided $t$-test rejects $H_0^{\text{C2}}$ at $p < 10^{-6}$ and $|d|_{\text{C2}} > 2$.

The verifier declares \textsc{Verified} only if \emph{both} the memorization criterion (C1) and the separation criterion (C2) pass, and \textsc{Rejected} otherwise. 
The two criteria target different null hypotheses and different risks. 
C1 asks whether the suspect $\mathcal{M}_B$ has actually memorized $\mathcal{W}$; C2 asks whether the owner $\mathcal{M}_A$ genuinely stands apart from public references that never saw $\mathcal{W}$. 
A \textsc{Verified} verdict therefore requires both criteria to satisfy
their respective decision rules. Both use the $p$-value cutoff
$10^{-6}$, and C2 additionally requires $|d|_{\text{C2}}>2$.

Because the probability that both tests reject at once cannot exceed the smaller of their two individual rejection probabilities, the type-I error of the joint decision is itself at most $10^{-6}$, and this inequality holds without assuming that the two tests are independent. 
The $10^{-6}$ thresholds here are the criterion-level significance applied to the C1 and C2 $p$-values; they should not be confused with the per-sample operating point $\alpha$ that fixes $\hat{q}_\alpha(x)$ inside C1. 

The conjunction is in fact conservative. 
In the false-accusation setting, $\mathcal{M}_B$ is neither derived from
$\mathcal{M}_A$ nor trained on $\mathcal{W}$. The null
$H_0^{\text{C1}}$ therefore holds, so the exact binomial test for C1
alone caps the false-positive rate at $10^{-6}$; C2 provides an
additional safeguard but is not needed for this bound.
Algorithm~\ref{alg:verification} states the complete procedure.

\begin{algorithm}[t]
\caption{Ownership Verification Protocol}
\label{alg:verification}
\begin{algorithmic}
\REQUIRE Models $\mathcal{M}_A, \mathcal{M}_B$, references $\{\mathcal{M}_{\text{ref}}^{(i)}\}_{i=1}^{|\mathcal{R}|}$, evidence set $\mathcal{W}$, calibrated quantile $\hat{q}_\alpha(\cdot)$ from Section~\ref{sec:score} with $\alpha = 10^{-3}$
\ENSURE $\textsc{Verified}$ or $\textsc{Rejected}$
\STATE $S_A \leftarrow \{s_{\mathcal{M}_A}(x) : x \in \mathcal{W}\}$; \quad $S_B \leftarrow \{s_{\mathcal{M}_B}(x) : x \in \mathcal{W}\}$
\STATE \textbf{C1 (memorization detection, binomial):}
\STATE \quad $k \leftarrow \sum_{x \in \mathcal{W}} \mathbf{1}\!\bigl[s_{\mathcal{M}_B}(x) \leq \exp(\hat{q}_\alpha(x)) - 1\bigr]$
\STATE \quad $p_1 \leftarrow \Pr\!\bigl[X \geq k \,\bigl|\, X \sim \mathrm{Binomial}(|\mathcal{W}|,\, \alpha)\bigr]$
\IF{$p_1 > 10^{-6}$} \STATE \textbf{return} $\textsc{Rejected}$ \ENDIF
\STATE \textbf{C2 (separation, paired):} \textbf{for each} reference $i \in \{1, \ldots, |\mathcal{R}|\}$:
\STATE \quad $S_{\text{ref}}^{(i)} \leftarrow \{s_{\mathcal{M}_{\text{ref}}^{(i)}}(x) : x \in \mathcal{W}\}$
\STATE \quad $\delta_{\text{C2}}^{(i)} \leftarrow S_A - S_{\text{ref}}^{(i)}$ \COMMENT{element-wise on $\mathcal{W}$}
\STATE \quad $p_2^{(i)} \leftarrow \text{OneSidedTTest}_{<}\bigl(\delta_{\text{C2}}^{(i)},\, \mu_0 = 0\bigr)$ \COMMENT{lower-tail: $H_1^{\text{C2}}:\mathbb{E}[\delta_{\text{C2}}]<0$}
\STATE \quad $|d|_{\text{C2}}^{(i)} \leftarrow \bigl|\,\text{mean}(\delta_{\text{C2}}^{(i)})\,\bigr| \big/ \text{std}(\delta_{\text{C2}}^{(i)})$
\IF{$\exists i: p_2^{(i)} > 10^{-6}$ \textbf{or} $|d|_{\text{C2}}^{(i)} < 2.0$}
    \STATE \textbf{return} $\textsc{Rejected}$
\ENDIF
\STATE \textbf{return} $\textsc{Verified}$
\end{algorithmic}
\end{algorithm}

\section{Experiments}
\label{sec:experiments}

\subsection{Experimental Setup}
\label{sec:setup}

We use two evaluation scenarios in our experiments:
(1) DDIM trained from scratch provides a controlled measurement of the protocol's quantitative properties.
(2) Stable Diffusion models with a fine-tuning dataset instantiate the production-scale threat surface.

\textbf{Datasets.}
We evaluate the DDIM model on four datasets spanning different resolutions and scales: CIFAR-10 and CIFAR-100 (both 32$\times$32, 50k training images with 5k reserved as the evidence set), STL-10 (96$\times$96, 5k training images with 1k as the evidence set), and CelebA (64$\times$64, 162k training images with 5k as the evidence set).

For the Stable Diffusion experiments, the owner's private training set consists of 1,000 images from COCO 2014 while the adversaries fine-tune on a disjoint COCO subset (also 1{,}000 images) plus a synthetic-image set.


\textbf{Baselines and Reference Models.}
\label{sec:setup_baselines}
For the baseline comparison in Section~\ref{sec:baseline_comparison}, we compare MiO against three diffusion-model IP protection methods: \textbf{WDM}~\cite{peng2023watermark}, a training-time watermarking method that learns a secondary watermark diffusion process; \textbf{Zhao et~al.}~\cite{zhao2023recipe}, which embeds StegaStamp fingerprints into training images before EDM training; and \textbf{CDI}~\cite{dubinski2024cdi}, an inference-time MIA-aggregation method. For Stable Diffusion, we additionally compare against \textbf{SleeperMark}~\cite{wang2025sleepermark}, a trigger-prompt-coupled, LoRA-resistant watermark method.

As public references for the verification protocol, we use pretrained checkpoints from HuggingFace: \texttt{google/ddpm-cifar10-32} serves as the reference for both CIFAR-10 and CIFAR-100 datasets, \texttt{google/ddpm-ema-bedroom-256} (resized) is for STL-10, and \texttt{google/ddpm-celebahq-256} is for CelebA.
These references are trained on disjoint data from our evidence sets and represent the null hypothesis in the two-point verification protocol.

\subsection{RQ1: How Reliable is MIA in Criteria 1 at Restricted FPR?}
\label{sec:fpr_control}

The calibrated quantile $\hat{q}_\alpha(x)$ turns the per-sample $t$-error into an FPR-$\alpha$ decision: we flag $x$ as a member when $\log(1 + s_{\mathcal{M}_A}(x)) \le \hat{q}_\alpha(x)$. 
For each owner family, we fit the Gaussian quantile regression ensemble
in Section~\ref{sec:score} and hold it fixed
across all suspect variants derived from that owner.
We then evaluate this indicator on held-out members and non-members for each owner, reporting ROC-AUC and TPR at the restricted FPR budgets $\alpha \in \{10^{-3},\, 10^{-4}\}$.
Because an ownership verification in practice must be held at a controlled false-accusation rate, we treat TPR at a fixed low FPR as the primary reliability metric.

Table~\ref{tab:mia_results} reports per-sample MIA performance for the four DDIM owner models and the three different Stable Diffusion v1.4 owner configurations. 
First, the AUC is uniformly high where every owner model attains at least 0.879 AUC.
Second, the TPR at the two target operating points varies substantially
across owner models.
Consequently, even at these restricted-FPR operating points, the detection rate remains far above the corresponding chance level, indicating that the $t$-error signal underlying the MIA is sufficiently reliable in the low-FPR regime.

These results answer RQ1 in the affirmative for the case we test: the $t$-error MIA retains usable true-positive rates at false-positive budgets as low as $10^{-4}$. 
Criterion~1 aggregates these per-sample decisions through the hit rate
$\hat{r}$ and tests the null $H_0^{\text{C1}}\!:\, r \le \alpha$.
Interpreting $\alpha$ as the binomial null rate requires the per-sample
decision rule to attain the corresponding false-positive rate.

\begin{table}[t]
\caption{Per-sample MIA performance under Gaussian QR calibration.}
\label{tab:mia_results}
\centering
\begin{small}
\begin{tabular}{@{}lccc@{}}
\toprule
Owner & AUC & \multicolumn{2}{c}{TPR} \\
\cmidrule(l){3-4}
 & & $\alpha{=}10^{-3}$ & $\alpha{=}10^{-4}$ \\
\midrule
\multicolumn{4}{@{}l}{\emph{DDIM (from scratch)}} \\
CIFAR-10  & \AUCCten   & \TPRtenthCten   & \TPRhundredthCten \\
CIFAR-100 & \AUCChund  & \TPRtenthChund  & \TPRhundredthChund \\
STL-10    & \AUCStl    & \TPRtenthStl    & \TPRhundredthStl \\
CelebA    & \AUCCeleb  & \TPRtenthCeleb  & \TPRhundredthCeleb \\
\midrule
\multicolumn{4}{@{}l}{\emph{Stable Diffusion v1.4}} \\
LoRA $r{=}64$  & \AUCLoraSm & \TPRtenthLoraSm & \TPRhundredthLoraSm \\
LoRA $r{=}256$ & \AUCLoraLg & \TPRtenthLoraLg & \TPRhundredthLoraLg \\
Full FT        & \AUCFullFT & \TPRtenthFullFT & \TPRhundredthFullFT \\
\bottomrule
\end{tabular}
\end{small}
\end{table}

\subsection{RQ2: Is Memorization Unique to the Owner?}
\label{sec:uniqueness}

\begin{table*}[t]
\caption{Paired $|d|_{\text{C2}}$ across DDIM and SD owners against three reference roles.}
\label{tab:main_results}
\centering
\begin{small}
\setlength{\tabcolsep}{6pt}
\begin{tabular}{@{}lccc@{}}
\toprule
Owner & Semantic-similar & Cross-domain & Random init \\
\midrule
\multicolumn{4}{@{}l}{\emph{DDIM (from scratch)}} \\
CIFAR-10  & \cmark~(23.9) & \cmark~(31.6) & \cmark~(45.1) \\
CIFAR-100 & \cmark~(16.3) & \cmark~(25.2) & \cmark~(44.1) \\
STL-10    & \cmark~(87.3) & \cmark~(22.2) & \cmark~(41.5) \\
CelebA    & \cmark~(20.5) & \cmark~(23.5) & \cmark~(63.0) \\
\midrule
\multicolumn{4}{@{}l}{\emph{Stable Diffusion v1.4}} \\
LoRA $r{=}64$  & \cmark~(2.54) & \markLoraSmCross~(\dCtwoLoraSmCross) & \markLoraSmRand~(\dCtwoLoraSmRand) \\
LoRA $r{=}256$ & \cmark~(2.61) & \markLoraLgCross~(\dCtwoLoraLgCross) & \markLoraLgRand~(\dCtwoLoraLgRand) \\
Full FT        & \cmark~(2.36) & \markFullFTCross~(\dCtwoFullFTCross) & \markFullFTRand~(\dCtwoFullFTRand) \\
\bottomrule
\end{tabular}
\end{small}
\end{table*}

In Criteria 2 of the two-point verification, we aim to separate the owner's model from public reference models by checking if the memorization on the private evidence set is only unique to the owner's model.
Table~\ref{tab:main_results} reports $|d|_{\text{C2}}$ (Eq.~\eqref{eq:paired_c2}) for seven owner models: four DDIM models on CIFAR-10, CIFAR-100, STL-10, CelebA trained from scratch and three Stable Diffusion v1.4 models fine-tuned on a 1{,}000-image COCO subset with LoRA at rank 64, LoRA at rank 256, full UNet fine-tuning.
Each model is evaluated against three reference models from different settings: semantic-similar, cross-domain, and random initialization. 

Semantic-similar means the reference model is trained/fine-tuned on data of the same type and distribution as the owner, but has never seen the owner's private evidence set $\mathcal{W}$.
Take DDIM as an example: for the CIFAR-10, CIFAR-100, and STL-10 owners, the semantic-similar reference models are uniformly \texttt{google/ddpm-cifar10-32} (trained on CIFAR-10, likewise 32$\times$32 natural images); 
for the CelebA owner, it is \texttt{google/ddpm-celebahq-256} (trained from the face domain).
Cross-domain means the reference model is trained/fine-tuned on data from an entirely different category. 
Trained on indoor bedroom scenes, LSUN's \texttt{google/ddpm-bedroom-256} is the cross-domain reference for CIFAR-10, CIFAR-100, and CelebA.
For STL-10, we adopt \texttt{google/ddpm-church-256} to avoid a domain-match confound with some of STL-10's classes since its training dataset is church images.
Random initialization reference model is a network at the dataset's native resolution with fully random weights and no training at all.

Stable Diffusion is analogous, except that the models are not trained from scratch but fine-tuned from SD v1.4. 
Its semantic-similar reference is therefore the un-fine-tuned base \texttt{CompVis/stable-diffusion-v1-4}: the owner model is fine-tuned from it on a 1{,}000-image COCO subset, so this reference shares the same SD v1.4 backbone as the owner yet was never fine-tuned on the owner's evidence set $\mathcal{W}$.
The cross-domain reference is \texttt{hakurei/waifu-diffusion-v1-3} (anime style), whose fine-tuning data is entirely disjoint from COCO.
The random-init reference is a randomly initialized UNet that reuses only the SD v1.4 VAE.

In the DDIM part of Table~\ref{tab:main_results}, all separations are significant. Across all 12 (owner, reference) cells, $|d|_{\text{C2}}$ ranges from a minimum of 16.3 to a maximum of 87.3, far beyond the threshold 2.0 in Criterion 2.
The reason is that the owner model $\mathcal{M}_A$ is trained entirely from scratch by us, so its memorization of $\mathcal{W}$ is way stronger than that of any reference model naturally.
For the latent-space SD owners, any reference that shares the SD v1.4 backbone reconstructs the same image to a comparable error, so the gap is compressed and $|d|_{\text{C2}}$ falls in a smaller range (e.g, between 2 and 3).
Only the randomly initialized UNet, which does not share the backbone, drives the separation up to about 19. 
Yet, across all 9 cells, $|d|_{\text{C2}}$ remains greater than 2, so the separation still holds under the threshold.
It is worth noting that no matter for DDIM or SD, the random-init reference gives the largest separation, further showing that the owner's memorization of its dedicated evidence set $\mathcal{W}$ is clearly stronger than that of any model not trained on it.

In summary, across both architecture families, all owners' models are separate from the corresponding reference models, reflecting the uniqueness of $\mathcal{M}_A$'s memorization on $\mathcal{W}$ that Criterion~2 demands.

\subsection{RQ3: Post-Theft Robustness}
\label{sec:theft}
In the real scenario, the adversary can apply various post-theft modifications to the stolen model, trying to evade IP detection methods.
Therefore, an ownership verification protocol is only useful when its signal survives realistic post-theft attacks. 
We test whether MIA memorization from Section~\ref{sec:t_error} can continue to flag members of the private evidence set $\mathcal{W}$ after the adversary mutates the stolen $\mathcal{M}_A$ into $\mathcal{M}_B$ via five attacks:
MMD-FT, SGD-FT, Gaussian perturbation with noise scale of $\sigma=10^{-3}, 10^{-2}$, and LoRA update.

\begin{table}[t]
\caption{Post-theft robustness of the MiO framework.}
\label{tab:robustness}
\centering
\small
\begin{tabular}{@{}lccc@{}}
\toprule
Attack on $\mathcal{M}_A$ & AUC & \multicolumn{2}{c}{TPR} \\
\cmidrule(l){3-4}
 & & $\alpha{=}10^{-3}$ & $\alpha{=}10^{-4}$ \\
\midrule
None (positive control: $\mathcal{M}_A$) & \AUCCten & \TPRtenthCten & \TPRhundredthCten \\
MMD-FT & \AUCMmd & \TPRtenthMmd & \TPRhundredthMmd \\
SGD-FT & \AUCSgd & \TPRtenthSgd & \TPRhundredthSgd \\
Noise $\sigma\!=\!10^{-3}$ & \AUCNoiseSm & \TPRtenthNoiseSm & \TPRhundredthNoiseSm \\
Noise $\sigma\!=\!10^{-2}$ & \AUCNoiseLg & \TPRtenthNoiseLg & \TPRhundredthNoiseLg \\
\bottomrule
\end{tabular}
\end{table}

We first describe the attack methods and experimental settings. 
For MMD-FT, we use 500 iterations with a cubic-polynomial kernel over a frozen CLIP ViT-B/32 to train MMD-FT, an aggressive method that reshapes the owner's model output distribution by matching CLIP features to the training distribution.
SGD-FT serves as a controlled version for MMD-FT, which is trained with exactly the same budget, but on clean data and under its own loss. 
Since both methods receive the same amount of training, any difference in their results must come from the type of attack rather than from unequal computing.
Gaussian weight noise has no training, but the isotropic perturbations of standard deviation $\sigma\in\{10^{-3},10^{-2}\}$ are added in place, standing in for a quantization or noise-injection adversary. 
For Stable Diffusion, we apply a 2{,}000-step cross-attention LoRA
fine-tune to each of the three owner configurations, varying the
adversary rank between 64 and 256 and the fine-tuning data between a
disjoint COCO subset and synthetic images, for a total of twelve
configurations.

The un-attacked positive control, i.e., the CIFAR-10 owner model of Table~\ref{tab:mia_results}, attains TPR$@10^{-4}$ of \TPRhundredthCten.
Each adversarial attack on $\mathcal{M}_B$ degrades this only partially.
Across the four weight-space attacks on the CIFAR-10 owner, the residual
AUC remains at least 0.976.
Table~\ref{tab:robustness} reports the four post-theft attacks evaluated
on the CIFAR-10 owner.
Overall, the ownership signal persists under MMD and SGD fine-tuning, weight noise at both magnitudes, and the latent-space LoRA attack, which retains a TPR@$10^{-4}$ of 0.614.

Furthermore, we tested the robustness of MiO against a stronger attack.
All attacks above start from the stolen model weights and modify them, while a more aggressive adversary could instead sample a large synthetic dataset from the stolen model and train a fresh model on it, which is a knowledge distillation-based attack.

The distillation attack is evaluated in Table~\ref{tab:distillation}.
50{,}000 synthetic CIFAR-10 images are sampled from the stolen model $\mathcal{M}_A$ with 10-step DDIM, and then a student model is trained from scratch for 100{,}000 iterations at batch 128. 
On the student model, the per-sample MIA collapses toward a random chance, where the student's hit rate on $\mathcal{W}$ stays at the $\alpha$ floor, so Criterion 1's binomial test fails to reject $H_0^{\text{C1}}$ and the protocol returns \textsc{Rejected}. 
Our protocol rejects the distilled student even though it was trained on
samples from $\mathcal{M}_A$ (Table~\ref{tab:distillation}).

\begin{table}[t]
\caption{Distillation attack performance on CIFAR-10}
\label{tab:distillation}
\centering
\small
\begin{tabular}{@{}lccc@{}}
\toprule
Model & AUC & \multicolumn{2}{c}{TPR} \\
\cmidrule(l){3-4}
 & & $\alpha{=}10^{-3}$ & $\alpha{=}10^{-4}$ \\
\midrule
Owner $\mathcal{M}_A$ & \AUCCten & \TPRtenthCten & \TPRhundredthCten \\
Distilled student & \AUCDistill & \TPRtenthDistill & \TPRhundredthDistill \\
\bottomrule
\end{tabular}
\end{table}

\subsection{RQ4: Controlled Baseline Comparison}
\label{sec:baseline_comparison}

Existing ownership verification methods are difficult to compare head-to-head on verification effectiveness, because each defines ``verification'' through a different statistical instrument, evaluated at a different operating point and sample-size regime.
For example, SSIM-based watermark similarity is used in WDM~\cite{peng2023watermark}, bit-accuracy on a fixed message is used in Zhao et~al.~\cite{zhao2023recipe} and SleeperMark~\cite{wang2025sleepermark}, and Welch's $t$-test on aggregated MIA features is used in CDI~\cite{dubinski2024cdi}.

Since these protocols share no common operating framework, we instead report each method under its own native detection score in Table~\ref{tab:cost_comparison}.
Under its own metric, every baseline attains near-perfect verification on a clean protected model: WDM extracts its watermark at an SSIM of 0.997, Zhao and SleeperMark recover their embedded message at 0.999 and 0.99 bit-accuracy, and CDI identifies its dataset with over 99\% confidence.
MiO attains a residual AUC of \AUCCten{} on the CIFAR-10 owner.
However, these scores cannot be collapsed into a single comparable number, such as TPR at a fixed FPR.

We therefore organize the baseline methods not by a single effectiveness score but along two \emph{cost} axes in Table~\ref{tab:cost_comparison}: 
(a)~whether the protocol modifies the owner model's weights, which risks degrading generation quality (FID);
and (b)~whether verification requires an auxiliary component coupled to the deployed model, as opposed to being performed offline on an unmodified model. 
MiO sits on the favorable end of both axes: 
(1) It never modifies the owner model's weights because the ownership signal is injected by a private dataset in a non-invasive way, 
and (2) its Gaussian quantile regression ensemble is constructed offline on the verifier side, so the ownership verification operates on an unmodified model without auxiliary components.
As shown in Table~\ref{tab:cost_comparison}, MiO requires no weight
modification, adds no overhead to the owner's training pipeline, and
introduces no post-training embedding step. WDM, Zhao et~al., and
SleeperMark instead rely on training-time or model-level embedding,
with the FID changes reported under their default configurations.

\begin{table}[t]
\caption{Computational cost breakdown on CIFAR-10 dataset.}
\label{tab:cost}
\centering
\begin{small}
\begin{tabular}{lcc}
\toprule
Component & Time & Parameters \\
\midrule
DDIM training (owner model) & $\sim$34 h & 30.5M \\
Gaussian QR ($\times 1$ model) & $\sim$2 h & 11.57M \\
Gaussian QR ensemble ($B{=}50$) & $\sim$100 h$^\dagger$ & 579M total \\
\midrule
\multicolumn{3}{l}{\textit{Inference (per verification query)}} \\
$t$-error scoring (5k samples) & $\sim$10 min & --- \\
Ensemble prediction & $<$1 min & --- \\
Two-point criteria & $<$1 s & --- \\
\midrule
\textbf{Total query time} & \textbf{$\sim$11 min} & --- \\
\bottomrule
\multicolumn{3}{l}{} \\
\end{tabular}
\end{small}
\end{table}

Specifically, CDI is the only baseline on the membership-based side amongst baselines, and the only prior method like MiO that requires zero weight modification or any deployment-time component.
However, it is a \emph{dataset-ownership} verification method rather than a \emph{model-ownership} verification method.

\begin{table*}[t]
\caption{Comparison across diffusion-model ownership-verification methods.}
\label{tab:cost_comparison}
\centering
\small
\setlength{\tabcolsep}{5pt}
\begin{tabular}{@{}lcccc@{}}
\toprule
Method & Modifies weights & Training overhead & $\Delta$FID & Detection score \\
\midrule
WDM & \cmark & High (extra WDP) & $+0.58$ (CIFAR-10) & SSIM $0.997$ \\
Zhao        & \cmark & High (encoder + DM)    & $+4.87$ (CIFAR-10) & Bit-acc $0.999$ \\
CDI         & \xmark & None                   & N/A                & $>\!99\%$ conf.\ (dataset) \\
SleeperMark & \cmark & Medium (LoRA backdoor) & $+0.48$ (SD v1.4)  & Bit-acc $\approx\!0.99$ \\
\midrule
\textbf{MiO (ours)} & \xmark & \textbf{None}   & \textbf{N/A}       & \textbf{AUC $\AUCCten{}$}  \\
\bottomrule
\end{tabular}
\end{table*}

Furthermore, we tested the training overhead of the MiO method in Table~\ref{tab:cost}.
Even though MiO does not need extra computation in embedding watermarks, building the Gaussian quantile regression ensemble offline on the verifier's side also takes a certain amount of cost. 
The first part of the cost is a one-time, upfront training cost.
Note that, although training the ensemble takes $\approx$100 V100-hours for $B{=}50$ when run serially on a single V100, it is parallelable across the $B{=}50$ models and can be reused for every query. 
The second part of the cost is the actual cost when an auditor judges the ownership. 
It covers $\approx$10 minutes to score the 5{,}000 evidence-set samples on a single V100, plus one minute for ensemble quantile prediction, and an immediate two-point verification.
Unlike watermarking-based methods, which spend much computation on re-training the diffusion model under a watermark objective, our MiO instead moves the verification to the verifier in an offline manner.

\subsection{Hyperparameters}
\label{sec:ablation}

There are a few hyperparameters in our MiO framework. 
In this section, we conduct an ablation study for each of them and test the best combination of them.

\textbf{$t$-error Aggregation Strategy.}
Table~\ref{tab:aggregation} compares four strategies for collapsing the $t$-error sequence into a single membership score: mean, median, and two lower quantiles Q10 and Q25.
We rank them by $|d|_{\text{C2}}$ as ~\ref{sec:criteria} defined, so a larger value is preferable.
Q25 yields the strongest separation ($|d|_{\text{C2}} = 23.93$), surpassing the mean ($18.2$), the median ($19.8$), and the more aggressive Q10 ($21.5$).
The same observation holds in the latent-space diffusion setting, so we adopt Q25 in this setting.

\begin{table}[t]
\caption{Aggregation ablation on CIFAR-10.}
\label{tab:aggregation}
\centering
\begin{small}
\begin{tabular}{lc}
\toprule
Aggregation & $|d|_{\text{C2}}$ \\
\midrule
Mean & $18.2$ \\
Q10 & $21.5$ \\
Q25 (ours) & $\mathbf{23.93}$ \\
Median & $19.8$ \\
\bottomrule
\end{tabular}
\end{small}
\end{table}

\textbf{The Size of Ensemble $B$.}
Varying the bagging ensemble of Gaussian quantile regression from $B=1$ to $100$, we measure the standard deviation of the predicted quantile $\hat{q}_\tau(x)$ over five runs on CIFAR-10 at FPR $0.1\%$. 
Table~\ref{tab:ensemble_ablation} shows that the prediction variance drops sharply from $B=1$ to $20$, keeps falling through $B=50$, and plateaus when beyond.
So, we adopt $B=50$ as a favorable stability and cost trade-off. 

\textbf{The Number of Sampled Timesteps.}
We vary $K$ from 10 to 100 and find out that $|d|$ plateaus once $K \geq 25$ and changes negligibly beyond $K = 50$. 
Therefore, we use the fixed $K = 50$ throughout.

\textbf{Gaussian QR vs.\ Pinball-loss QR.}
There are two mainstream approaches to quantile regression: Gaussian QR and pinball-loss QR.
Pinball-loss QR estimates a target quantile directly: it trains a regressor under the pinball-loss to calibrate at FPR $\alpha$, an asymmetric loss whose minimizer is the conditional $\alpha$-quantile of the score.
At the extreme low FPR operating points we report, $\alpha \in \{10^{-3},10^{-4}\}$, the two approaches yield comparable TPR.
Gaussian QR's advantage is flexibility: one fitted model returns the threshold $\hat{q}_\alpha(x)$ for any $\alpha$ in closed form via $\Phi^{-1}(\alpha)$, whereas pinball-loss QR requires a separate model per quantile level.
In our experiments, each model is evaluated across datasets at several FPR levels rather than one fixed operating point, so we need a calibrator that supplies every $\alpha$ from a single training, which leads to the adoption of Gaussian quantile regression.

\textbf{Latent vs.\ Pixel Space in Stable Diffusion.}
As established in Section~\ref{sec:problem}, the reconstruction error $e_t$ for Stable Diffusion can be computed in either latent or pixel space.
On SD~v1.4, latent-space scoring with per-image caption conditioning yields a higher AUC $= 0.9956$, compared to $0.959$ for pixel-space scoring since the VAE decode step dilutes the membership signal in pixel space.
So all SD experiments in our settings use the latent space configuration.

\begin{table}[t]
\caption{Effect of ensemble size $B$ on quantile prediction stability }
\label{tab:ensemble_ablation}
\centering
\begin{small}
\begin{tabular}{lcc}
\toprule
Ensemble $B$ & Std($\hat{q}_\tau$) & TPR @ FPR 0.1\% \\
\midrule
1 & 0.142 & 0.83 $\pm$ 0.05 \\
10 & 0.068 & 0.87 $\pm$ 0.03 \\
20 & 0.041 & 0.88 $\pm$ 0.03 \\
50 (ours) & 0.019 & 0.89 $\pm$ 0.03 \\
100 & 0.016 & 0.89 $\pm$ 0.02 \\
\bottomrule
\end{tabular}
\end{small}
\end{table}

\section{Conclusion}
\label{sec:conclusion}
This work studies ownership verification for diffusion models using a
protocol that does not modify the owner model or its sampling pipeline.
To overcome the limitations of existing watermarking methods, which often inject artificial signals into model weights or outputs and may degrade utility, this paper proposes Membership is Ownership (MiO), a method that relies on the private member set as a non-invasive watermark.
MiO leverages the model’s natural memorization behavior on the private member evidence dataset and formulates ownership verification as a population-level hypothesis test. 
Specifically, MiO combines membership-based attribution on the private
evidence set with statistical separation from public reference models.
Each criterion uses a $p$-value cutoff of $10^{-6}$.
Experimental results show that MiO achieves stable and reliable ownership verification with negligible impact on model utility, and remains robust under post-theft fine-tuning and weight perturbation attacks. 
Overall, MiO provides a non-invasive approach to ownership verification
for diffusion models.

\section*{Acknowledgment}
This work was supported in part by the U.S. National Science Foundation
(NSF) under Grant Nos.~2416872, 2315596, 2244219, 2146497, 2551417,
2548961, 2343619, 2429960, 2434899, and 2548041.

\bibliographystyle{IEEEtran}
\bibliography{references}

\end{document}